\documentclass{jaa}
\usepackage{natbib}
\usepackage{graphicx}
\usepackage{pdflscape}
\usepackage{amsmath}	
\usepackage{amssymb}
\usepackage{multirow}
\usepackage{multicol}
\usepackage{hyperref}

\begin{document}\sloppy

\title{Protoplanetary disks around young stellar and substellar objects in the $\sigma$ Orionis cluster}


\author{Belinda Damian\textsuperscript{1,*}, Jessy Jose\textsuperscript{2}, Beth Biller\textsuperscript{3,4} and KT Paul\textsuperscript{1}}
\affilOne{\textsuperscript{1}Department of Physics and Electronics, CHRIST (Deemed to be University), Hosur Road, Bengaluru 560029, India\\}
\affilTwo{\textsuperscript{2}Indian Institute of Science Education and Research (IISER) Tirupati, Rami Reddy Nagar, Karakambadi Road, Mangalam (P.O.), Tirupati 517507, India\\}
\affilThree{\textsuperscript{3}SUPA, Institute for Astronomy, University of Edinburgh, Blackford Hill, Edinburgh EH9 3HJ, UK\\}
\affilFour{\textsuperscript{4}Centre for Exoplanet Science, University of Edinburgh, Edinburgh, UK}


\twocolumn[{

\maketitle

\corres{belinda.damian@res.christuniversity.in, jessyvjose1@gmail.com}


\begin{abstract}
Understanding the evolution and dissipation of protoplanetary disks are crucial in star and planet formation studies. We report the protoplanetary disk population in the nearby young $\sigma$ Orionis cluster (d$\sim$408 pc; age$\sim$1.8 Myr) and analyse the disk properties such as dependence on stellar mass and disk evolution. We utilise the comprehensive census of 170 spectroscopic members of the region refined using astrometry from Gaia DR3 for a wide mass range of $\sim$19-0.004 M$_\odot$. Using the near infrared (2MASS) and mid infrared (WISE) photometry  we classify the sources based on the spectral index into class I, class II, flat spectrum and class III young stellar objects. The frequency of sources hosting a disk with stellar mass $<$2 M$_\odot$ in this region is 41$\pm$7\% which is consistent with the disk fraction estimated in previous studies. We see that there is no significant dependence of disk fraction on stellar mass among T Tauri stars ($<$2 M$_\odot$), but we propose rapid disk depletion around higher mass stars ($>$2 M$_\odot$). Furthermore we find the lowest mass of a disk bearing object to be $\sim$ 20 M$_\mathrm{Jup}$ and the pronounced disk fraction among the brown dwarf population hints at the formation scenario that brown dwarfs form similar to low-mass stars.
\end{abstract}

\keywords{Star forming regions---Protoplanetary disks---Young stellar objects.}

}]


\doinum{12.3456/s78910-011-012-3}
\artcitid{\#\#\#\#}
\volnum{000}
\year{0000}
\pgrange{1--}
\setcounter{page}{1}
\lp{1}

\section{Introduction}
\label{sec:intro}
In the star formation scenario circumstellar disks form around protostars as a consequence of the conservation of angular momentum (\citealt{andre2000,williams2011}). The presence of these disks have been evidenced through observational studies of the accretion signatures and infrared excess emission due to the dusty disk. As the system evolves the disk dissipates through i) accretion on to the central protostar channeled by the magnetic field lines from the inner edge of the disk to the stellar surface (see reviews by \citealt{bouvier2007,hartmann2016}),  ii) accumulation into planets (see review by \citet{miotello2022}, iii) photoevaporation due to the radiation from the protostar (see reviews by \citealt{frank2014,pascucci2022}) and iv) external photoevaporation due to nearby massive stars (\citealt{hartmann2009,winter2018}). The optically thick inner disk present during the early phases of disk evolution produces excess over the stellar photosphere that can be detected by studying the young stellar objects (YSOs) in the near infrared (NIR) and mid infrared (MIR) wavelengths \citep{hartigan1995}. 

The lifetime of the disk depends on the disk dissipation timescale. This has a direct effect on the planet formation since the circumstellar disks are the birth sites for planets and provide raw materials for its formation. Hence they impact when and what type of planets can form (\citealt{meyer2007,currie2009}). Previous studies have shown that the timescale for the disk evolution is dependent on the stellar mass (\citealt{lada1995,calvet2004}). The disk frequency around massive stars ($>$1 M$_\odot$) have been observationally reported to be lower than that in low-mass stars which implies a rapid disk dispersal in early type stars (\citealt{lada2006,dahm2007,hernandez2007int}).  

Due to the dependence of disk fraction on stellar mass, the time available for the formation of planets around early type stars is low. However since the disk mass is relative to the stellar mass, there is a higher possibility to encounter giant planets in the vicinity of massive stars \citep{johnson2010}. Hence quantifying the disk frequency as a function of stellar mass and age is essential to understand the disk lifetime and in turn constrain the planet formation theories (\citealt{andrews2020,rilinger2021}). 

The $\sigma$ Orionis cluster located near the Horsehead nebula below the Orion belt in the Ori OB1b association is one of the well studied regions in the Orion complex. Due to its proximity (d $\sim$ 408 pc; \citet{damian2023}) and low extinction (A$_\mathrm{v}$ $\sim$ 0.4mag; \citet{damian2023}) we have a strong census of the sources in the region including the very low-mass objects. Also due to its youth ($\sim$2 Myr; \citet{damian2023}) this is an ideal target to study the disk properties of young stars in their primordial stages of evolution. Having a robust catalogue of spectroscopically confirmed members in the core of the region we study the disk properties such as disk frequency and its dependence on stellar mass and age.

The paper is organised as follows: section~\ref{sec:data} describes the data used in this work, section~\ref{sec:mem_sel} discusses the selection of members of the region and section~\ref{sec:phy_par} presents the estimation of the physical parameters of the members. In section~\ref{sec:yso_class} we classify the members based on the IR SED slope into different evolutionary stages of YSOs and in section~\ref{sec:results} we estimate the disk fraction and discuss its dependence on stellar mass as well as the disk evolution. Section~\ref{sec:conclusion} summarises the results of this work.

\section{Data}
\label{sec:data}
The primary data set of the members of the $\sigma$ Orionis cluster is taken from \citet{damian2023}. In \citet{damian2023}, NIR photometry from the Wide-field InfraRed Camera (WIRCam) on the Canada-France-Hawaii Telescope (CFHT) \citep{puget2004} in the J-, H- and W-bands for a survey area of $\sim$ 21 x 21 arcmin centered at RA = 84.700$^{\circ}$ and Dec = -2.568$^{\circ}$ is used. Along with this,  to classify the YSOs and to characterize the disk bearing members of the region we use the NIR and MIR data from the Two Micron All Sky Survey (2MASS) \citep{cutri2003} and Spitzer \citep{spitzer2021}/ Wide-field Infrared Survey Explorer (WISE) \citep{cutri2012} respectively. The CFHT WIRCam data is used as the base catalogue and we find counterparts for those sources with a cross-match radius of 1 arcsec in 2MASS (J, H and Ks), Spitzer (IRAC 3.6, 4.5, 5.8 and 8 $\mathrm{\mu}$m) and WISE (W1 (3.4 $\mu$m), W2 (4.6 $\mu$m), W3 (12 $\mu$m) and W4 (22 $\mu$m)).

\section{Membership selection}
\label{sec:mem_sel}
We use the catalogue of 170 spectroscopically confirmed members of the $\sigma$ Orionis region from \citet{damian2023}. The selection and compilation of the members is briefly described below. The members compiled in \citet{damian2023} comprise of newly confirmed  sources as well as known members in literature. Firstly we identified the brown dwarf candidates using the CFHT WIRCam data based on the reddening insensitive index (Q) which is defined as follows,

\begin{equation}
      Q = (J-W) + e(H-W)
\end{equation}

\noindent where J, H and W are the magnitudes from the corresponding filters and e is the ratio of extinction in each of these bands. The Q index is defined to identify brown dwarfs in star forming regions and distinguish them from foreground and background field stars independent of reddening (refer \citet{allers2020}). The value of Q is scaled such that Q=0 corresponds to M0 type objects, which do not show significant water absorption feature and decreasing values of Q corresponds to objects of later spectral types (for instance Q=-0.6 relates to M6 type object ($\sim$0.08 M$_{\odot}$)), which show prominent water absorption in their atmosphere. The selection criteria Q$<$-(0.6 + 3$\sigma_Q$) is used to select the candidate very low-mass stars and brown dwarfs in the region where the uncertainty in Q ($\sigma_Q$) is estimated by propagating the errors in J, H and W (see \citet{jose2020} for details). 

Secondly the candidates were selected using the Gaia DR3 \citep{gaia2022} photometric and astrometric data. We incorporate the following astrometric quality conditions: renormalized unit weight error (RUWE) $<$ 1.4, which filters the sources with unreliable astrometric solutions (\citealt{das2023,kordopatis2023,stoop2023,manara2021,fabricius2021}) and $\sigma_\pi$/$\pi$ $<$0.1 where $\pi$ is parallax and $\sigma_\pi$ is the uncertainty in parallax (\citealt{esplin2022,penoyre2022}). Sources satisfying the above conditions and located along the pre-main sequence (PMS) branch in the Gaia$_\mathrm{G}$ - J vs absolute Gaia$_\mathrm{G}$ colour magnitude diagram (CMD) with parallax and proper motion consistent with that of the cluster are selected as candidate members. These astrometric conditions are applied to known members of the region in literature as well, to filter sources which have astrometry inconsistent with that of the cluster (refer \citet{damian2023} for more details).

Together we identified 28 low-mass stars and brown dwarf candidates and obtained their spectra with the 3.2m NASA Infrared Telescope Facility (IRTF) SpeX \citep{rayner2003}. The observations were carried out in the low-resolution prism mode (R$\sim$150) covering the wavelength range 0.7-2.5 $\mu$m. Along with these newly identified members, the previously known members of the $\sigma$ Orionis cluster within the CFHT survey area ($\sim$ 21 x 21 arcmin) were combined comprising of 170 spectroscopically confirmed members including the massive O9.5 star ($\sigma$ Ori) down to the planetary mass sources. The complete catalog of members is adopted as is from \citet{damian2023} for this work. The data is complete down to $\sim$ 0.004 M$_\odot$ in all the three CFHT WIRCam J-, H- and W-bands.

\section{Physical parameters of the members}
\label{sec:phy_par}
The parameters of the members of the region like spectral type, effective temperature (T$_\mathrm{eff}$), bolometric luminosity and mass estimated in \citet{damian2023} are incorporated in this work. The spectral types for the previously known members are adopted from literature and for the newly identified low-mass sources in \citet{damian2023} are obtained through comparison of the IRTF NIR spectra with standard spectral templates. The spectral types for the 170 members range from O9.5 for the massive star $\sigma$ Ori at the center of the cluster to L5 for the lowest planetary mass source. T$_\mathrm{eff}$ is obtained from the spectral type using the appropriate relations from \citet{pecaut2013} (for O9.5-M5 type), \citet{herczeg2014} (for M5.5-M6.5 type) and \citet{filippazzo2015} (for M7-L5 type). Then the bolometric luminosity for the massive O, B and A type sources are incorporated based on their spectral type from the extended table of \citet{pecaut2013}. Whereas for the other spectral type sources it is derived from the bolometric magnitude incorporating the bolometric correction (BC) given by the following equation,

\begin{equation}
    log(L_\mathrm{bol}/L_{\odot}) = -\frac{(M_\mathrm{bol} - M_\mathrm{bol,\odot})}{2.5}
\end{equation}

\noindent where M$_\mathrm{bol}$  =  m$_\mathrm{J}$ - 5log(d) + 5 - A$_\mathrm{J}$ + BC$_\mathrm{J}$ and the distance to the region is estimated to be 408$\pm$8 pc based on Gaia DR3 parallax; the mean extinction towards the cluster $A_\mathrm{v}$=0.4 mag and the solar bolometric magnitude M$_\mathrm{bol,\odot}$=4.73 mag. Utilising these T$_\mathrm{eff}$ and luminosity estimates, the mass of the sources are determined from the HR-diagram using the isochrones and evolutionary models of PMS stars from \citet{baraffe2015} (for mass $>$ 0.01 M$_{\odot}$) and the models from BT-Cond \citep{baraffe2003} (for mass $<$ 0.01 M$_{\odot}$). The average age of the cluster was estimated as 1.8$\pm$1 Myr. The member list from \citet{damian2023} along with their physical parameters are given in ~\ref{sec:appendix}.

\section{Classification of YSOs using spectral index}
\label{sec:yso_class}
We classified the different evolutionary stages of the YSOs following the methodology described in \citet{koenig2015}. This can be quantified using the slope of the spectral energy distribution from NIR to MIR wavelengths \citep{lada1987}. \citet{greene1994} classified the YSOs based on the value of the infrared spectral index ($\alpha$) which reflects the slope of the SED, into four stages of evolution: Class I, Class II, Flat spectrum and Class III. The slope $\alpha$ is expressed as,

\begin{equation}
    \alpha = \frac{d ~log ~(\lambda F_{\lambda})}{d ~log ~\lambda}
\end{equation}

\noindent In the above equation the wavelength ($\lambda$) and flux density (F$_{\lambda}$) are calculated using the zero-point for the respective filters taken from the VOSA filter profile service {\footnote{\url{http://svo2.cab.inta-csic.es/theory/fps/}}}.
Here we compute $\alpha$ with the dereddened 2MASS K$_\mathrm{s}$ ($\lambda$ = 2.1 $\mu$m) and WISE W4 ($\lambda$ = 22 $\mu$m) bands whenever possible. To deredden the data we use the mean extinction towards the cluster $A_\mathrm{v}$=0.4 mag and the extinction relation from \citet{cardelli1989} and \citet{koenig2014} for the 2MASS and WISE magnitudes respectively. We apply the condition that the uncertainty in K$_\mathrm{s}$ should be non-null with the SNR in W4 $\geq$ 5 and the reduced $\chi^{2}$ of the W4 profile-fit photometry measurement ($\chi^{2}_\mathrm{W4}$) between 0.3-1.7 \citep{koenig2015}. These conditions are satisfied by 34 sources whose SED slope is estimated using the K$_\mathrm{s}$ and W4 wavelengths. When a source does not comply with the criteria for W4, then we use the W3 ($\lambda$ = 11.6 $\mu$m) band with the same condition for SNR and $\chi^{2}_\mathrm{W3}$ which is the case for 74 sources. When both W4 and W3 photometry does not satisfy the condition, we then resort to W2 ($\lambda$ = 4.6 $\mu$m) filter as for 35 sources. For these 35 sources estimating $\alpha$ with H- instead of K-band photometry does not result in a significant change in the YSO classification. In the case when K$_\mathrm{s}$ data is unavailable or with null photometric error then the slope value is calculated between W1 (SNR $\geq$ 3) and W4 (SNR $\geq$ 5; 0.3 $<$ $\chi^{2}_\mathrm{W4}$ $<$ 1.7) \citep{koenig2015}. Twenty five sources either do not have WISE data or 2MASS data or their photometry do not meet the quality requirements. Therefore, we obtain the spectral index for 145 members of the region (see~\ref{sec:appendix}) and classify the sources as Class I ($\alpha$ $\geq$ 0.3), Flat spectrum (0.3 $>$ $\alpha$ $\geq$ -0.3), Class II (-0.3 $>$ $\alpha$ $\geq$ -1.6) and Class III ($\alpha$ $<$ -1.6) (\citealt{greene1994,koenig2015}).

Table~\ref{tab:yso_class} summarises the YSO class of the members. Among the 145 sources that are classified into various YSO classes, the majority of the sources are identified as class III objects. Further analysis on the disk fraction and its significance are presented in section~\ref{sec:results} below.  

\begin{table}[htb]
\tabularfont
    \caption{Spectral index based classification of the members of the $\sigma$ Orionis cluster.}
     \label{tab:yso_class}
    \begin{tabular}{lcc}
    \topline
     Class & $\alpha$ & Number of sources  \\\midline
    I     & $\geq$0.3 & 2\\
    Flat spectrum & $<$0.3 and $\geq$-0.3 & 3\\
    II & $<$-0.3 and $\geq$-1.6 & 54\\
    III & $<$-1.6 & 86\\
    \hline
    \end{tabular}
\end{table}

\begin{figure*}[t!]
\centering\includegraphics[width=\textwidth,scale=0.7]{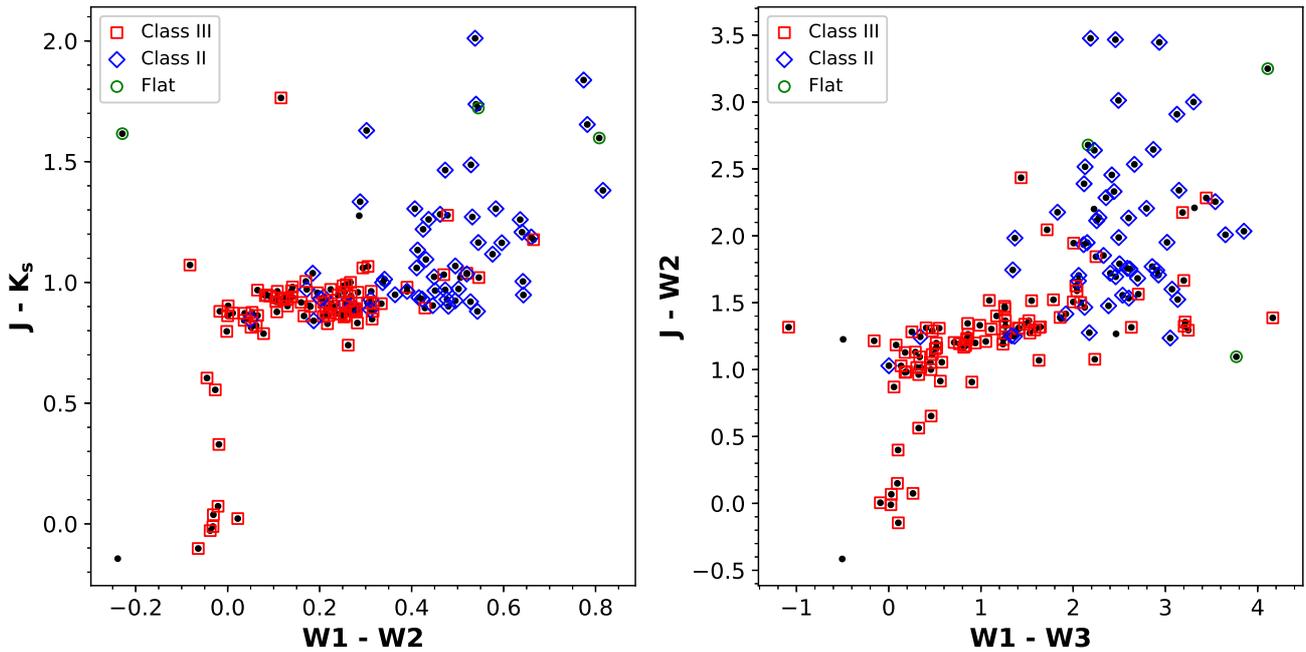}
\caption{\textit{Left}: WISE W1-W2 vs 2MASS J-K$_\mathrm{s}$ colour-colour diagram of the members of the $\sigma$ Orionis cluster (black dots). The excess and non-excess sources classified using the SED slope ($\alpha$) are marked with different colours as presented in the legend. \textit{Right}: WISE W1-W3 vs 2MASS J-WISE W2 colour-colour diagram with the same colour scheme as the left figure.  }
\label{fig:ccd}
\end{figure*}

\begin{figure}
    \centering
    \includegraphics[width=\columnwidth]{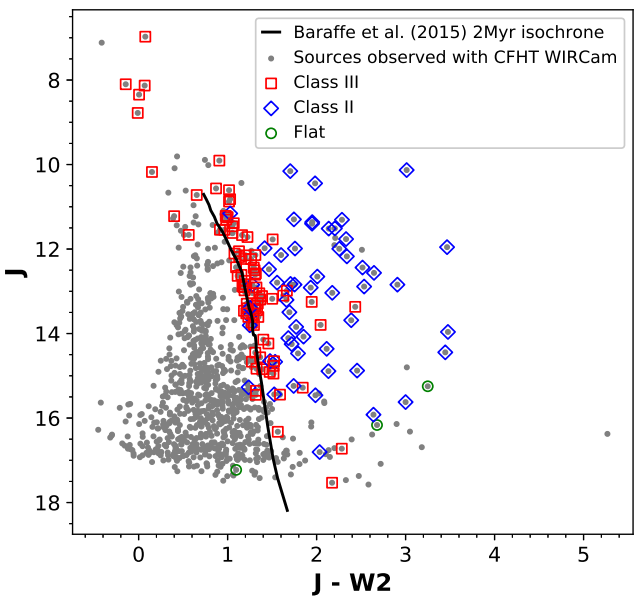}
    \caption{2MASS J - WISE W2 vs J color-magnitude diagram. The markers represent the different types of YSOs as given in the legend. The black line shows the 2 Myr isochrone from \citet{baraffe2015}.}
    \label{fig:cmd}
\end{figure}

In figure~\ref{fig:ccd} we show the colour-colour diagram with 2MASS and WISE highlighting the distribution of the disk and diskless sources. We see a distinct colour excess for the Class II sources indicating the presence of a circumstellar disk around them. The sources with negative MIR colour excess (in W1-W2) are mainly the early type stars of O to A spectral class. Likewise in figure~\ref{fig:cmd} we show the 2MASS J - WISE W2 CMD for the sources observed with the WIRCam from \citet{damian2023} as mentioned in section~\ref{sec:data}. Here we see that the Class III objects are aligned along the PMS branch consistent with the 2 Myr isochrone \citep{baraffe2015}, whereas the sources hosting disk are to the right due to the NIR excess emission from its disk. 

\begin{figure*}
\centering\includegraphics[trim=0 20 0 0,clip,scale=0.65]{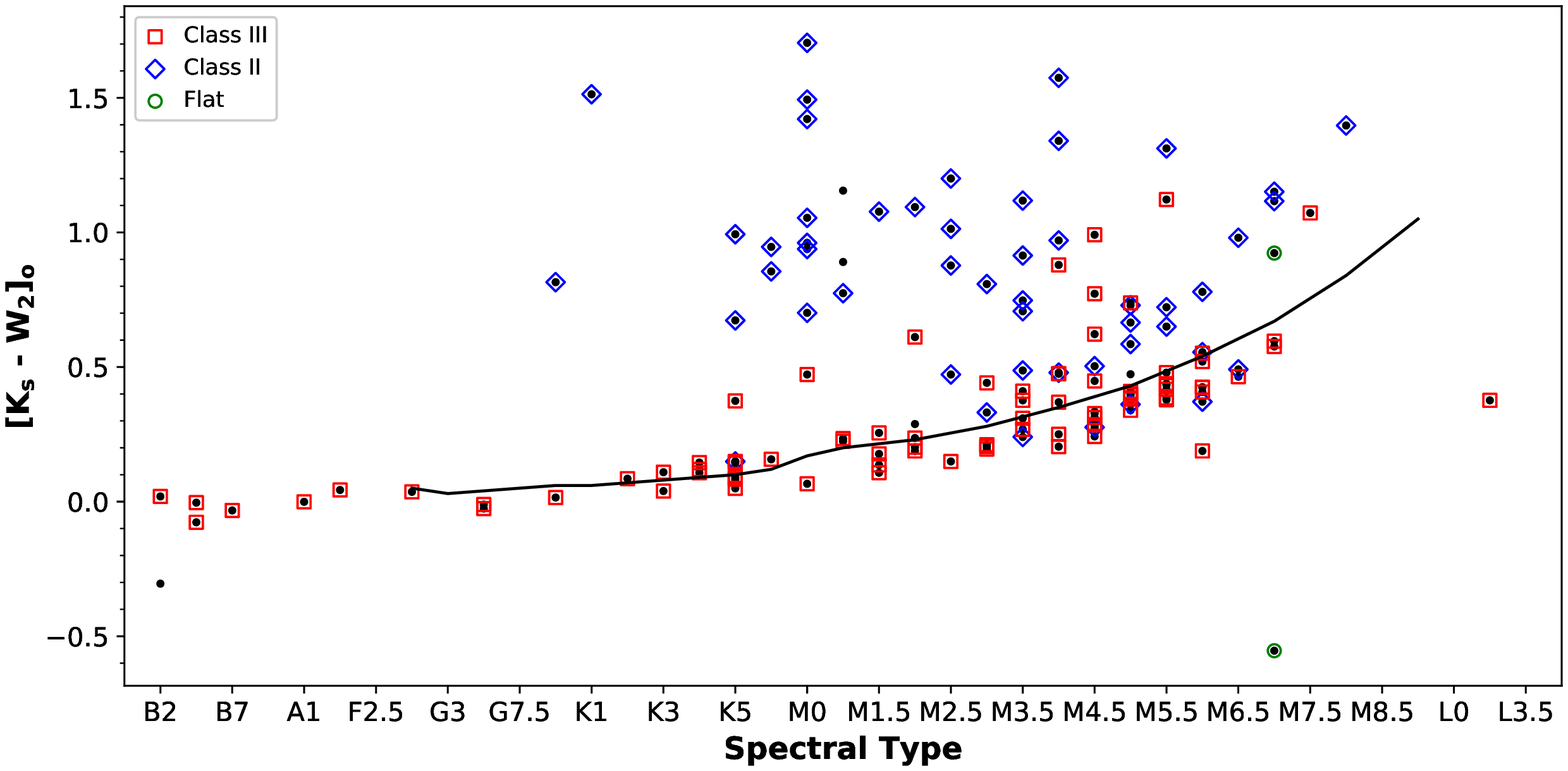}
\centering\includegraphics[scale=0.65]{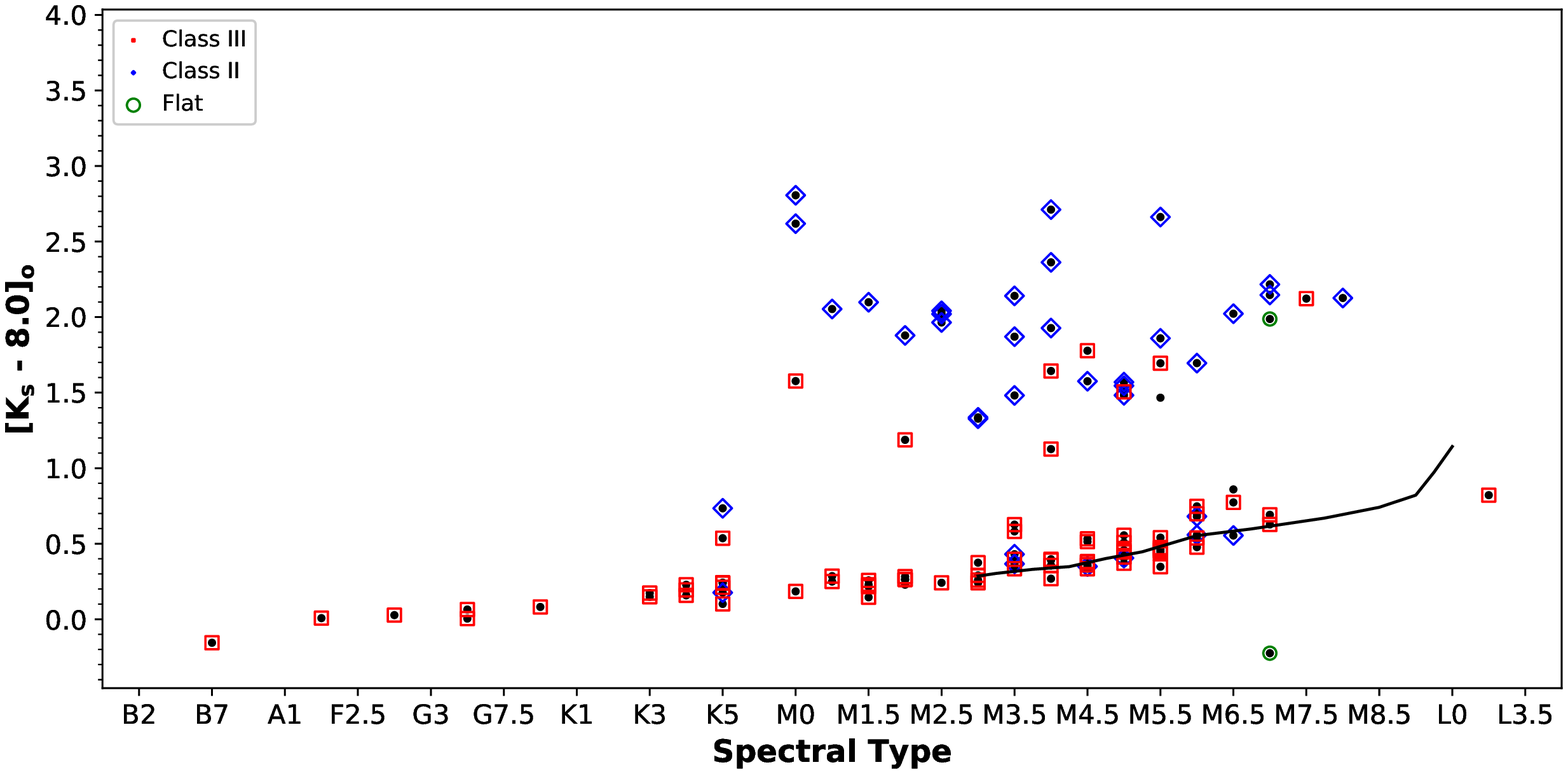}
\caption{Spectral type as a function of extinction corrected K$_\mathrm{s}$-W2 and K$_\mathrm{s}$-[8.0] colour. The markers represent the different types of YSOs as given in the legend. The black line indicates the sequence of the intrinsic photospheric colour of young stars taken from \citet{pecaut2013} (top panel) and \citet{esplin2017} (bottom panel). }
\label{fig:spt_vs_colr}
\end{figure*}

Figure~\ref{fig:spt_vs_colr} shows the relation between the spectral type of the sources and the extinction corrected colour. To deredden the 2MASS, WISE and Spitzer data we adopt the mean extinction to the cluster A$_\mathrm{v}$ = 0.4 mag \citep{damian2023} and the reddening relation from \citet{cardelli1989}, \citet{koenig2014} and \citet{wang2019} respectively. The continuous line marks the trend of the intrinsic photospheric colour as a function of spectral type from \citet{pecaut2013} table 6 for K$_\mathrm{s}$-W2 colour and from  \citet{esplin2017} for  K$_\mathrm{s}$-[8.0] colour. We compare the colours of the sources classified as Class II with the typical photoshperic colours of YSOs for a given spectral type and we see that the dereddened  K$_\mathrm{s}$-[8.0] colour shows an excess of $>$ 1 mag. Likewise in the top panel of figure~\ref{fig:spt_vs_colr} the Class II sources exhibit an excess in the dereddened K$_\mathrm{s}$-W2 colour than the corresponding photospheric sequence indicating that the excess is likely due to the presence of a circumstellar disk in the system. We also note from the distribution of disk bearing (Class II) and diskless sources (Class III) in the figure that the color excess decreases towards the late M type objects relative to the excess seen in the early more massive sources possibly caused by the more massive disks around them.

\section{Results and discussion}
\label{sec:results}

\subsection{Disk fraction}
\label{sec:disk_frac}
We estimate the disk fraction of the $\sigma$ Orionis cluster which essentially gives the percentage of disk bearing sources among the members of the region. Here we consider the sources identified as Class II, Class I and Flat spectrum as disk bearing sources and the disk fraction is calculated as the number of disk bearing sources divided by the total number of sources. Since there are no disk sources among the massive members we only consider sources with mass $<$ 2 M$_\odot$ for the estimate. This yields a disk frequency of 41\% $\pm$ 7$\%$ (error is estimated from Poisson uncertainty). Previous disk studies on this region based on the Spitzer IRAC colour excess have quoted similar results and some of them are reported below. \citet{hernandez2007} reported that the disk fraction depends on the stellar mass ranging from ~$\sim$10\% for Herbig AeBe stars to $\sim$35\% for T Tauri stars. \citet{luhman2008} estimated the disk frequency for low-mass stars as $\sim$40\% and for the brown dwarfs as $\sim$60\%. Later \citet{pena2012} derived a similar result of $\sim$40\% for the very low-mass stars, brown dwarfs and planetary-mass objects. These previous studies on the $\sigma$ Orionis cluster have reported a disk fraction which is consistent with our current estimate based on a comprehensive membership of spectroscopic members  down to the planetary mass regime in the core of the region.

\subsection{Dependence of disk fraction on stellar mass} 
\label{sec:mass_dependence}
\begin{figure*}
\centering\includegraphics[width=\textwidth,scale=0.7]{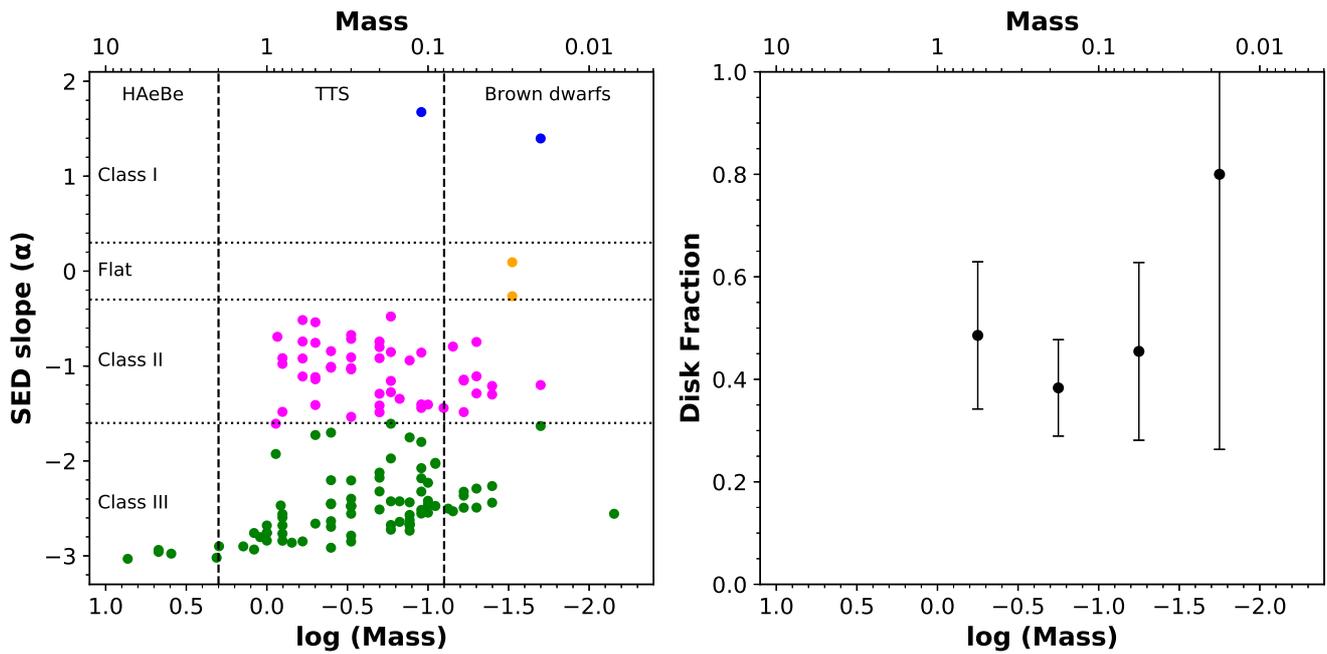}
\caption{\textit{Left}: Logarithmic stellar mass as a function of the SED slope estimated in section~\ref{sec:yso_class}. The vertical dashed lines denote the mass range of the Herbig AeBe stars, T Tauri stars and brown dwarfs. The horizontal dotted lines mark the range of the $\alpha$ values for each YSO class. \textit{Right}: Disk fraction for every logarithmic mass bin of size 0.5dex. The error bars denote the Poisson error at each point. }
\label{fig:mass_vs_alpha}
\end{figure*}

With the estimated disk fraction we obtained its correlation and dependence on the stellar mass as shown in figure~\ref{fig:mass_vs_alpha}. The confirmed members of the region were grouped together based on their mass into Herbig AeBe stars ($>$ 2 M$_\odot$; \citet{guzman2021}), T Tauri stars (2-0.08 M$_\odot$; \citet{furlan2006}) and brown dwarfs ($<$ 0.08 M$_\odot$; \citet{luhman2012}). The left panel of figure~\ref{fig:mass_vs_alpha} shows the distribution of the various classes of objects in relation to their stellar mass and in the right panel we chart the fraction of disk bearing sources emitting IR excess for every logarithmic bin of 0.5 dex. 

We see that there is no significant variation of disk fraction with respect to mass and the majority of the disk sources consist of T Tauri stars with a scarcity of excess sources towards the high mass end (mainly above 2 M$_\odot$). Although the disk fraction seems to increase in the lowest mass bin (i.e. the planetary mass objects) however, this must be treated with caution due to the high uncertainty in the disk fraction estimate and also since we do not classify 25 YSOs (refer section~\ref{sec:yso_class}) this could cause incompleteness in the data towards the low-mass end. The low disk frequency for stars with mass $>$1 M$_\odot$ could be a signature of rapid disk evolution of the high mass stars whereas for the low-mass stars with higher excess this can be attributed to longer disk lifetime.

A similar case was reported in the nearby ($\sim$ 300 pc) young ($\sim$ 1-2 Myr) cluster NGC 1333 by \citet{yuhan2018}. Their scarce population of intermediate mass stars does not favour a robust comparison of disk fraction with the low-mass population. However they estimate the disk fraction for the low-mass stars (0.1-1.5 M$_\odot$) to be consistent, lacking any significant variation with respect to mass. \citet{yuhan2018} also analysed the dependence of disk fraction on stellar mass for the young star forming region IC 348 ($\sim$ 3 Myr; $\sim$ 320 pc). They found that the high mass region (2.2-5 M$_\odot$) hosts the lowest disk frequency and the disc frequency increases towards the low-mass stars (0.25-1.5 M$_\odot$).

\citet{ribas2015} carried out a detailed study to test the influence of stellar mass on the protoplanetary disk evolution using a large sample of spectroscopically confirmed YSOs ($\sim$1400) in 22 nearby associations ($<$500 pc) spread across different ages of $\sim$ 1-100 Myr. They grouped the sample into two based on stellar mass with a boundary mass of 2 M$_\odot$. They found robust evidence for dependence of disk lifetime on stellar mass where the high mass stars were found to disperse their disk rapidly, twice as quickly as low mass stars. This is due to the lower efficiency of the mechanisms responsible for removing circumstellar gas and dust around low mass stars leading to longer timescales for planet formation around such hosts.This is in agreement with our observations of disk dispersal in the $\sigma$ Orionis cluster between T Tauri and Herbig AeBe stars.


\subsection{Analysis on NIR excess}
\label{sec:disk_excess}
\begin{figure}
    \includegraphics[width=\columnwidth,scale=0.55]{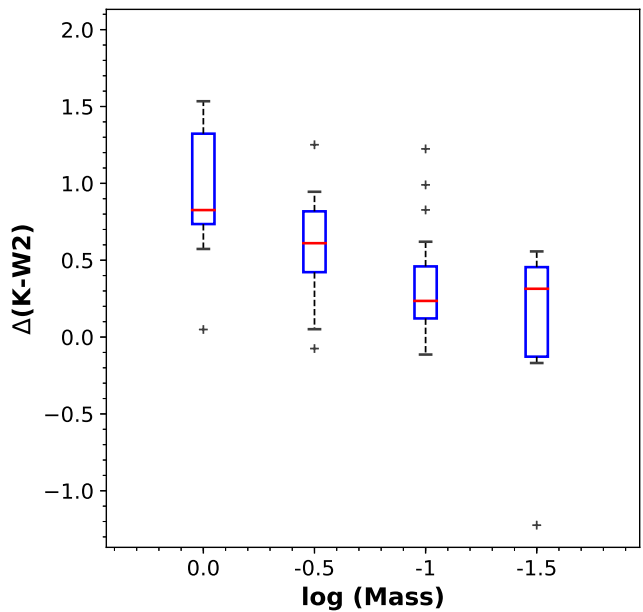}
    \caption{Box and whisker plot between mass and color excess of YSOs with disk in the $\sigma$ Orionis cluster. The box marks the lower (Q1) and upper (Q3) quartile range and the red horizontal line denotes the median for a bin of size 0.5 dex  (width of the box has no significance here). The whisker extends from the 5th to the 95th percentile and the outliers are marked with the plus symbol.}
    \label{fig:mass_colorexc}
\end{figure}

The three major contributing factors for the  observed color of the YSOs are the photospheric color, the circumstellar excess and interstellar reddening. To determine the NIR excess due to the presence of disk, we follow the definition proposed in \citet{hillenbrand1998} as given in the below equation. We use the 2MASS K-band and WISE W2 magnitudes to obtain the excess.

\begin{equation}
    \Delta(K-W2) = (K-W2)_{o} - (K-W2)_{phot}
\end{equation}

\noindent where (K-W2)$_{o}$ is the extinction corrected observed color and (K-W2)$_{phot}$ is the intrinsic photospheric color. The extinction A$_\mathrm{v}$ = 0.4 mag is adopted from \citet{damian2023} and we use the reddening relation from \citet{cardelli1989} and \citet{koenig2014} to obtain extinction in K- and W2-bands respectively. The (K-W2)$_{phot}$ as a function of spectral type is taken from \citet{pecaut2013}. In figure~\ref{fig:mass_colorexc} we present the distribution of NIR excess for the disk bearing sources (class I, class II and flat spectrum sources) identified in section~\ref{sec:yso_class}. Although we see a slight decline in the NIR excess as we move towards the very low-mass end, we cannot conclusively deduce the dependence of disk excess on the stellar mass due to two caveats -- scarcity of high mass sources harbouring disk mainly above 1 M$_\odot$ and the low number density of such sources in the region resulting in a high uncertainty in the estimated NIR color excess. \citet{yuhan2018} analysed the NIR excess of sources hosting disk around them in the Orion A molecular cloud using 2MASS K-band and Spitzer 4.5$\mu$m color. They found that the excess decreases for stars in between the mass range of 0.36-0.42 M$_\odot$.


\subsection{Disk evolution}
\label{sec:disk_evo}
The gas and dust in the protoplanetary disks are the building blocks for planets. Therefore the frequency and dispersal of disks around stars of different ages is important to constrain planet formation theories. 

\begin{figure}[htb]
\includegraphics[width=\columnwidth,scale=0.7]{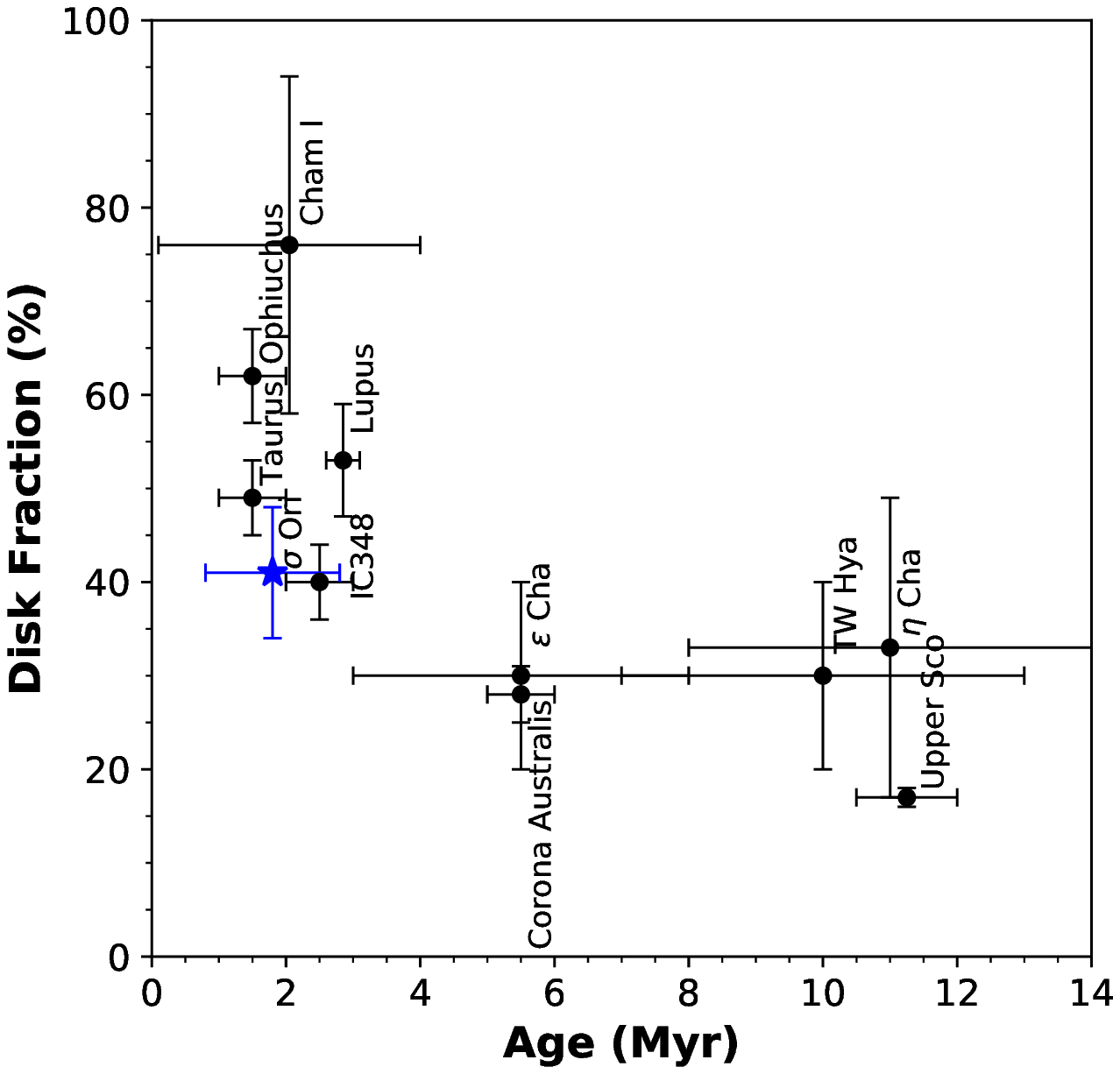}
\caption{Age as a function of disk fraction in different star forming regions. The data points for Ophiuchus, Taurus, Cham I, IC348, Lupus, $\epsilon$ Cha, Corona Australis, TW Hydra, $\eta$ Cha and Upper Scorpius are taken from \citet{michel2021} with their respective errors. The disk fraction of the $\sigma$ Orionis cluster estimated in this work is shown with the blue star and the error bar denotes the Poisson error.}
\label{fig:age_vs_df}
\end{figure}

Figure~\ref{fig:age_vs_df} shows the relation between the disk fraction and the age of various young star forming regions from literature along with the relation for the $\sigma$ Orionis cluster reported in this work. In \citet{damian2023} the mean age of the cluster is computed as the average age of the confirmed members of the region from the HR-diagram as 1.8 $\pm$ 1 Myr. The disk fraction and age of other star forming regions are adopted from \citet{michel2021} (table 1) where the disk fraction is estimated based on the the SED slope ($\alpha_\mathrm{Lada}$) between 2-22 $\mu$m (i.e. 2MASS K-WISE4) which is similar to the wavelengths we have utilised for $\sigma$ Orionis to estimate the disk fraction and the $\alpha$ range for YSO classification. The age of the star forming regions range between 1-12 Myr with decreasing disk frequency for older regions. We see that our estimate for the $\sigma$ Orionis cluster of 41\% at 1.8 Myr matches with the trend of other star forming regions (\citealt{das2021,gupta2021,panwar2017}). 


Among the various mechanisms contributing to the dispersion of protoplanetary disks, the two external factors that affect disk lifetime are external photoevaporation due to the ultraviolet (UV) flux from massive stars in the vicinity and due to tidal truncation caused by close encounters. \citet{winter2018} quantified the threshold for a system to be under the influence of these two mechanisms by measuring the far-UV (FUV) flux driving the external photoevaporation and the stellar number density attributed to the tidal truncation for star forming regions under varying local environments. They derived a threshold for the number density to be $>$10$^4$ pc$^{-3}$ for a period of 3 Myr above which the effect of tidal truncation is significant in disk dispersion. Additionally for the host star of $\sim$1 M$_\odot$ with a massive disk around it $\sim$0.1 M$_\odot$ the minimum FUV flux required to dissipate the disk within 3 Myr is G$_{o}$ $>$ 3000 (where G$_{o}$ is the FUV flux of the interstellar). The number density and the FUV flux in the $\sigma$ Orionis cluster is below the minimum values reported in \citet{winter2018} for the system to be influenced by external photoevaporation and tidal truncation (refer figure 3 of \cite{winter2018}). On the other hand, \citet{ansdell2017} surveyed the gas and dust in disks around YSOs in the region using ALMA and found a decline in the disk dust mass at smaller separations from the central massive star ($\sigma$ Ori). They claim that the external photoevaporation influences disk evolution in the region. However as the latter study probes mainly the outer part of the disk, which could explain the discrepancy. Therefore we cannot conclusively rule out the effect of external photoevaporation on disk evolution in this region and our robust sample of members can facilitate in refining further works in this regard.

\subsection{Disks around brown dwarfs and planetary mass objects}
\label{sec:disks_bd}
The formation scenario of brown dwarfs and planetary mass objects ($<$0.08 M$_\odot$) is still an open question. Various mechanisms have been proposed to explain the formation of these substellar objects such as turbulent fragmentation - the compression of gas in the molecular cloud forms cores which collapses into low mass objects (\citealt{padoan2002,andre2012}); dynamical ejection - the core is ejected out of a multiple system before it could accrete enough mass to start hydrogen burning \citep{reipurth2001}; photoerosion - the outer layers of a prestellar core are eroded by nearby massive stars before it accretes onto the central protostar \citep{whitworth2004}; disk fragmentation - gravitational fragmentation of the disk material \citep{stamatellos2011} (see review by \citet{luhman2012}). Similar to newly formed stars, brown dwarfs have been observed to host a disk around them (\citealt{muench2001,natta2002}). 

In young star forming regions it has been observed that the frequency of disks harbored by brown dwarfs are comparable with the disk fraction among stars \citep{luhman2005}. The presence of disk around these substellar objects are indicative of their formation mechanism which supports the theory that brown dwarfs form like stars. Additionally observational studies have reported the presence of planetary mass companions hosted by brown dwarfs, analogues to planets around stars (\citealt{han2013, shvartzvald2017}). Protoplanetary disks contain the raw materials and are the birth sites of planets. Therefore understanding the evolution of the disks is crucial in constraining the brown dwarf and planet formation theories \citep{rilinger2021}.   

Another interesting question to investigate is whether these planetary mass objects can host their own planetary systems. Detecting signatures of disk around these objects would unravel their nature and shed light on their formation mechanism. However it is challenging to observe these ultra cool objects due to their low luminosity and distinguish them from field contaminants. This requires deep imaging of significantly large areas of young star forming regions for a statistically robust sample followed by spectroscopic confirmation of youth. A recent study by \citet{scholz2023} using Spitzer photometry in 3.6 and 4.5 $\mu$m reported the lowest mass of an object harboring a disk in the NGC 1333 region to be $\sim$ 10 M$_\mathrm{Jup}$. They also reviewed various other nearby star forming regions and found the results to be consistent. This may indicate that $\sim$ 10 M$_\mathrm{Jup}$ could be the lower limit for the opacity limited fragmentation to form like stars. With our deep photometry and W-band technique of identifying these ultra cool objects, along with the IR excess analysis using WISE data, we find that for the central region of the $\sigma$ Orionis cluster, the lowest mass objects with a disk are $\sim$ 20 M$_\mathrm{Jup}$ (WBIS\_053854.9-024033.8 and WBIS\_053900.3-023705.9). WBIS\_053900.3-023705.9 (UGCS J053900.29-023705.7) and WBIS\_053854.9-024033.8 (2MASS J05385492-0240337) are classified as class I and class II YSOs respectively (refer table~\ref{tab:appendix}). 

In the $\sigma$ Orionis cluster, we see a pronounced population of protoplanetary disks around brown dwarfs (refer figure~\ref{fig:mass_vs_alpha}). Although the MIR data towards the very low mass end is incomplete, we see that among the 25 substellar sources ($<$0.08 M$_\odot$) that were classified in section~\ref{sec:yso_class}, 14 sources harbour disks (class I + class II + flat spectrum) which gives the upper limit of the disk fraction in this regime. 
Given that we have used a robust sample of spectroscopically confirmed brown dwarfs in the region despite the lack of a large sample, the significant disk fraction in the substellar mass range cannot be overlooked. This non-negligible disk fraction among the brown dwarf population provides clues to its possible origin similar to that of low-mass stars formed via turbulent fragmentation of the molecular cloud.


\section{Conclusions}
\label{sec:conclusion}

Utilizing the comprehensive catalog of spectroscopic members in the nearby young $\sigma$ Orionis cluster from \citet{damian2023}, we have analysed the protoplanetary disk properties of the YSOs. Our members have also been refined based on astrometry from Gaia DR3.  We have used the 2MASS and WISE NIR photometry to classify the members based on the infrared spectral index into the four YSO classes - class I, class II, flat spectrum and class III. We then verified the presence of the disk around the YSOs by comparing their color excess with the intrinsic photospheric color for a given spectral type as well as their distribution in the NIR and MIR color-color diagrams. We see that the class II sources show a  distinct color excess indicating that the excess is due to the presence of a circumstellar disk around the source. The disk frequency of the $\sigma$ Orionis cluster was found to be 41$\pm$7\% and this correlates well with the age of the cluster when compared with other star forming regions.  

We also analysed the dependence of disk fraction on stellar mass which showed that there is no significant variation of disk frequency with respect to the mass among T Tauri stars ($<$2 M$_\odot$) and the population of sources hosting disk decreases among high mass stars ($>$2 M$_\odot$) suggesting rapid disk dispersal. Furthermore we measured the disk color excess between the 2.2$\mu$m K-band and 4.6 $\mu$m W2-band and although the color excess seems to decrease with decreasing mass we require a large sample across a wider mass range to robustly correlate both. Our robust sample of spectroscopic brown dwarfs in the region are found to host a significant population of protoplanetary disks around them resulting in a non-negligible disk fraction for the substellar objects. We find that the lowest mass of objects with a disk are $\sim$ 20 M$_\mathrm{Jup}$. This supports the formation mechanism which proposes the formation of brown dwarfs similar to low-mass stars. 


\appendix

\section{Details of the members in  the $\sigma$ Orionis cluster}
\label{sec:appendix}

\begin{landscape}
\begin{table}
\tabularfont
\caption{The catalogue of spectroscopically confirmed members of the $\sigma$ Orionis cluster. The spectral type, effective temperature (T$_\mathrm{eff}$), bolometric luminosity and mass are taken from \citet{damian2023}. Column 6 and 7 are the YSO classifications reported for our sources in \citet{hernandez2007} and \citet{koenig2015} respectively. Column 8 gives the SED slope ($\alpha$) estimated in section~\ref{sec:yso_class} and the corresponding YSO class is given in column 9.}
\label{tab:appendix} 
\begin{tabular}{lcccccccc}
\topline
Object ID & SpT & T$_\mathrm{eff}$ & log(L$_\mathrm{bol}$/L$_\mathrm{\odot}$) & Mass & Class (H07)  & Class (K15) & $\alpha$ & Class (This work)\\
& & (K) & & (M$_\mathrm{\odot}$) & & & & \\ \midline
WBIS\_053900.3-023705.9	&	M8	&	2539	&	-2.47	&	0.02	&	-	&	-	&	1.40	&	I	\\
WBIS\_053832.1-023243.1	&	M5	&	2880	&	-0.94	&	0.11	&	II	&	-	&	1.68	&	I	\\
WBIS\_053848.0-022714.2	&	M0	&	3770	&	0.42	&	0.50	&	II	&	II	&	-1.41	&	II	\\
WBIS\_053925.2-023822.0	&	K7	&	3970	&	-0.02	&	0.60	&	II	&	II	&	-0.92	&	II	\\
WBIS\_053831.4-023633.8	&	M3.5	&	3260	&	-0.46	&	0.20	&	II	&	II	&	-0.74	&	II	\\
WBIS\_053834.1-023637.5	&	M4	&	3160	&	-0.40	&	0.17	&	II	&	II	&	-1.16	&	II	\\
WBIS\_053827.3-024509.6	&	M0	&	3770	&	-0.30	&	0.50	&	II	&	II	&	-0.76	&	II	\\
WBIS\_053845.4-024159.6	&	M1	&	3630	&	-0.33	&	0.40	&	II	&	II	&	-1.02	&	II	\\
WBIS\_053831.6-023514.9	&	M0	&	3770	&	-0.12	&	0.50	&	II	&	II	&	-1.14	&	II \\
WBIS\_053929.3-022721.0	&	M0	&	3770	&	-0.68	&	0.60	&	II	&	II	&	-0.52	&	II		\\
WBIS\_053827.5-023504.2	&	M3.5	&	3260	&	-0.77	&	0.20	&	II	&	II	&	-0.80	&	II	\\
WBIS\_053918.8-023053.2	&	K7	&	3970	&	-0.05	&	0.60	&	II	&	II	&	-1.11	&	II	\\
WBIS\_053907.6-023239.1	&	K5	&	4140	&	0.01	&	0.80	&	TD	&	TD	&	-0.92	&	II		\\
WBIS\_053908.8-023111.5	&	M3	&	3360	&	-0.81	&	0.30	&	CII	&	II	&	-1.04	&	II		\\
WBIS\_053859.2-023351.4	&	M2.5	&	3425	&	-0.82	&	0.30	&	II	&	II	&	-0.67	&	II	\\
WBIS\_053838.2-023638.5	&	K5	&	4140	&	0.06	&	0.80	&	III	&	-	&	-1.48	&	II		\\
WBIS\_053904.6-024149.2	&	M0	&	3770	&	-0.82	&	0.60	&	II	&	II	&	-0.74	&	II	\\
WBIS\_053823.3-022534.6	&	M2	&	3490	&	-1.06	&	0.40	&	II	&	II	&	-1.01	&	II		\\
WBIS\_053833.7-024414.3	&	K1	&	4920	&	0.57	&	0.86	&	II	&	II	&	-0.69	&	II	\\
WBIS\_053854.9-022858.3	&	M4.5	&	3020	&	-1.12	&	0.13	&	III	&	-	&	-0.94	&	II		\\
WBIS\_053806.7-023022.7	&	M1.5	&	3560	&	-0.26	&	0.40	&	II	&	II	&	-0.84	&	II		\\
WBIS\_053905.2-023300.5	&	-	&	-	&	-	&	-	&	III	&	-	&	-0.96	&	II	\\
WBIS\_053813.2-022608.8	&	M4.5	&	3020	&	-0.71	&	0.15	&	II	&	II	&	-1.35	&	II	\\
WBIS\_053817.8-024050.1	&	M5	&	2880	&	-0.91	&	0.11	&	II	&	II	&	-0.86	&	II	\\
WBIS\_053816.1-023804.9	&	M3.5	&	3260	&	-0.98	&	0.20	&	III	&	-	&	-0.92	&	II	\\
WBIS\_053911.5-023106.5	&	M0	&	3770	&	-0.32	&	0.50	&	II	&	II	&	-0.54	&	II		\\
WBIS\_053841.6-023028.9	&	M3	&	3360	&	-0.80	&	0.30	&	II	&	-	&	-0.71	&	II		\\
WBIS\_053820.5-023408.9	&	M4	&	3160	&	-0.79	&	0.17	&	II	&	II	&	-0.48	&	II		\\
WBIS\_053848.2-024400.8	&	-	&	-	&	-	&	-	&	II	&	II	&	-0.61	&	II	\\
WBIS\_053808.3-023556.2	&	M2.5	&	3425	&	-0.61	&	0.30	&	II	&	II	&	-0.91	&	II	\\
WBIS\_053835.9-024351.1	&	K0	&	5030	&	0.46	&	0.88	&	EV	&	TD	&	-1.61	&	II	\\
WBIS\_053903.0-024127.1	&	M2.5	&	3425	&	-0.67	&	0.30	&	II	&	II	&	-1.02	&	II		\\
WBIS\_053912.9-022453.5	&	M6	&	2860	&	-2.34	&	0.05	&	-	&	-	&	-0.75	&	II	\\
WBIS\_053850.6-024242.9	&	-	&	-	&	-	&	-	&	II	&	II	&	-1.18	&	II	\\
WBIS\_053838.9-022801.7	&	M5	&	2880	&	-1.75	&	0.06	&	III	&	-	&	-1.14	&	II	\\
\hline
\end{tabular}
\end{table}
\end{landscape}

\begin{landscape}
\begin{table}[htb]
\tabularfont
\begin{tabular}{lcccccccc}
\topline
Object ID & SpT & T$_\mathrm{eff}$ & log(L$_\mathrm{bol}$/L$_\mathrm{\odot}$) & Mass & Class (H07)  & Class (K15) & $\alpha$ & Class (This work) \\
& & (K) & & (M$_\mathrm{\odot}$) & & & & \\ \midline
WBIS\_053840.3-023018.5	&	M0	&	3770	&	-0.12	&	0.50	&	II	&	II	&	-1.12	&	II\\
WBIS\_053826.8-023846.1	&	M3.5	&	3260	&	-1.20	&	0.20	&	II	&	II	&	-1.29	&	II		\\
WBIS\_053915.1-024047.6	&	M3.5	&	3260	&	-1.37	&	0.20	&	III	&	-	&	-1.42	&	II	\\
WBIS\_053839.8-023220.3	&	M7	&	2683	&	-1.56	&	0.04	&	II	&	-	&	-1.21	&	II	\\
WBIS\_053855.4-024120.8	&	M5.5	&	2920	&	-1.72	&	0.07	&	II	&	-	&	-0.79	&	II	\\
WBIS\_053901.9-023502.8	&	M4	&	3160	&	-1.40	&	0.17	&	II	&	II	&	-0.85	&	II		\\
WBIS\_053835.4-022522.2	&	M6	&	2860	&	-1.50	&	0.06	&	III	&	-	&	-1.49	&	II	\\
WBIS\_053825.4-024241.2	&	M7	&	2683	&	-1.59	&	0.04	&	II	&	-	&	-1.30	&	II		\\
WBIS\_053913.1-023750.9	&	M6	&	2860	&	-1.70	&	0.06	&	II	&	-	&	-1.15	&	II		\\
WBIS\_053839.0-024532.1	&	M2.5	&	3425	&	-0.77	&	0.30	&	II	&	II	&	-1.54	&	II		\\
WBIS\_053840.5-023327.6	&	M5	&	2880	&	-0.83	&	0.11	&	II	&	II	&	-1.44	&	II		\\
WBIS\_053849.3-022357.5	&	M3.5	&	3260	&	-1.28	&	0.20	&	II	&	II	&	-1.49	&	II		\\
WBIS\_053848.1-022853.6	&	M5.5	&	2920	&	-1.39	&	0.08	&	II	&	-	&	-1.44	&	II		\\
WBIS\_053832.4-022957.3	&	M6.5	&	2815	&	-1.79	&	0.05	&	III	&	-	&	-1.11	&	II	\\
WBIS\_053926.3-022837.7	&	M5	&	2880	&	-1.02	&	0.10	&	II	&	II	&	-1.41	&	II	\\
WBIS\_053926.9-023656.1	&	M6.5	&	2815	&	-1.80	&	0.05	&	II	&	-	&	-1.29	&	II		\\
WBIS\_053833.9-024507.8	&	-	&	-	&	-	&	-	&	II	&	II	&	-1.42	&	II	\\
WBIS\_053843.9-023706.9	&	M5.5	&	2920	&	-0.85	&	0.11	&	II	&	II	&	-1.40	&	II		\\
WBIS\_053844.2-024019.7	&	K5	&	4140	&	-0.02	&	0.80	&	II	&	II	&	-0.98	&	II	\\
WBIS\_053847.2-023436.9	&	M4	&	3160	&	-0.74	&	0.17	&	II	&	II	&	-1.28	&	II		\\
WBIS\_053854.9-024033.8	&	M8	&	2539	&	-1.90	&	0.02	&	II	&	-	&	-1.20	&	II		\\
WBIS\_053807.1-024321.1	&	M7	&	2683	&	-2.43	&	0.03	&	-	&	-	&	0.09	&	flat	\\
WBIS\_053925.6-023843.7	&	-	&	-	&	-	&	-	&	-	&	II	&	0.22	&	flat		\\
WBIS\_053812.6-023637.7	&	M7	&	2683	&	-1.95	&	0.03	&	II	&	-	&	-0.26	&	flat		\\
WBIS\_053923.4-024057.5	&	M7.5	&	2611	&	-2.20	&	0.02	&	-	&	-	&	-1.63	&	III		\\
WBIS\_053918.1-022928.5	&	K0	&	5030	&	0.35	&	0.88	&	DD	&	-	&	-1.92	&	III		\\
WBIS\_053847.5-022712.0	&	M5	&	2880	&	-0.65	&	0.13	&	EV	&	-	&	-2.66	&	III	\\
WBIS\_053834.8-023415.7	&	A1	&	9300	&	1.49	&	2.05	&	III	&	-	&	-3.02	&	III		\\
WBIS\_053838.5-023455.0	&	K2	&	4760	&	0.64	&	0.82	&	EV	&	-	&	-2.47	&	III		\\
WBIS\_053834.2-023416.0	&	B7	&	14000	&	2.48	&	3.92	&	III	&	-	&	-2.98	&	III		\\
WBIS\_053901.5-023856.4	&	B5	&	15700	&	2.77	&	4.70	&	III	&	-	&	-2.96	&	III		\\
WBIS\_053841.3-023722.6	&	K5	&	4140	&	-0.06	&	0.80	&	III	&	-	&	-2.68	&	III		\\
WBIS\_053844.2-023233.6	&	K4	&	4330	&	0.20	&	1.00	&	III	&	-	&	-2.76	&	III		\\
WBIS\_053832.8-023539.2	&	M0	&	3770	&	-0.14	&	0.50	&	III	&	-	&	-2.66	&	III		\\
WBIS\_053853.4-023323.0	&	K3	&	4550	&	0.34	&	1.20	&	III	&	-	&	-2.76	&	III		\\
WBIS\_053925.6-023404.2	&	M5.5	&	2920	&	-0.93	&	0.11	&	III	&	-	&	-2.18	&	III		\\
\hline
\end{tabular}
\end{table}
\end{landscape}

\begin{landscape}
\begin{table}[htb]
\tabularfont
\begin{tabular}{lcccccccc}
\topline
Object ID & SpT & T$_\mathrm{eff}$ & log(L$_\mathrm{bol}$/L$_\mathrm{\odot}$) & Mass & Class (H07)  & Class (K15) & $\alpha$ & Class (This work) \\
& & (K) & & (M$_\mathrm{\odot}$) & & & & \\ \midline
WBIS\_053835.9-023043.3	&	K4	&	4330	&	0.05	&	1.00	&	III	&	-	&	-2.68	&	III		\\
WBIS\_053843.6-023325.4	&	M1	&	3630	&	-0.22	&	0.4	&	III	&	-	&	-2.46	&	III		\\
WBIS\_053921.0-023033.5	&	-	&	-	&	-	&	-	&	III	&	-	&	-2.06	&	III		\\
WBIS\_053842.3-023714.8	&	M0	&	3770	&	-0.23	&	0.50	&	II	&	II	&	-1.73	&	III		\\
WBIS\_053849.2-024125.1	&	M6	&	2860	&	-0.32	&	0.13	&	III	&	-	&	-2.66	&	III		\\
WBIS\_053900.5-023939.0	&	G3.5	&	5680	&	0.05	&	1.10	&	-	&	-	&	-2.80	&	III\\
WBIS\_053926.8-024258.3	&	-	&	-	&	-	&	-	&	EV	&	II	&	-1.73	&	III		\\
WBIS\_053839.7-024019.7	&	M3.5	&	3260	&	-1.06	&	0.20	&	III	&	-	&	-2.18	&	III	\\
WBIS\_053902.8-022955.8	&	M4	&	3160	&	-0.74	&	0.17	&	III	&	-	&	-2.67	&	III	\\
WBIS\_053853.2-024352.6	&	M1	&	3630	&	-0.43	&	0.40	&	III	&	-	&	-2.63	&	III	\\
WBIS\_053924.4-023401.3	&	M2	&	3490	&	-0.75	&	0.30	&	III	&	-	&	-2.21	&	III		\\
WBIS\_053905.4-023230.3	&	K5	&	4140	&	-0.09	&	0.80	&	III	&	-	&	-2.77	&	III		\\
WBIS\_053807.8-023130.7	&	K3	&	4550	&	0.36	&	1.20	&	III	&	-	&	-2.93	&	III		\\
WBIS\_053901.2-023638.8	&	-	&	-	&	-	&	-	&	III	&	-	&	-2.36	&	III	\\
WBIS\_053833.4-023617.6	&	M2.5	&	3425	&	-0.39	&	0.30	&	III	&	-	&	-2.85	&	III		\\
WBIS\_053911.6-023602.8	&	K5	&	4140	&	-0.12	&	0.80	&	III	&	-	&	-2.60	&	III	\\
WBIS\_053827.5-024332.5	&	A2	&	8800	&	1.38	&	1.98	&	III	&	-	&	-2.90	&	III	\\
WBIS\_053932.6-023944.0	&	K5	&	4140	&	0.20	&	0.70	&	III	&	-	&	-2.86	&	III		\\
WBIS\_053836.5-023312.7	&	B5	&	15700	&	2.77	&	4.70	&	III	&	-	&	-2.94	&	III		\\
WBIS\_053917.2-022543.3	&	M1.5	&	3560	&	-0.67	&	0.40	&	III	&	-	&	-2.45	&	III		\\
WBIS\_053835.5-023151.6	&	K4	&	4330	&	0.03	&	1.00	&	III	&	-	&	-2.84	&	III		\\
WBIS\_053843.3-023200.8	&	M5	&	2880	&	-0.71	&	0.13	&	III	&	-	&	-2.44	&	III		\\
WBIS\_053850.4-022647.7	&	M3.5	&	3260	&	-0.68	&	0.20	&	EV	&	-	&	-2.32	&	III		\\
WBIS\_053912.3-023006.4	&	M5	&	2880	&	-0.77	&	0.13	&	III	&	-	&	-2.57	&	III	\\
WBIS\_053847.7-023037.4	&	M5.5	&	2920	&	-1.02	&	0.11	&	III	&	-	&	-2.32	&	III		\\
WBIS\_053827.7-024300.9	&	M3	&	3360	&	-0.45	&	0.30	&	III	&	-	&	-2.79	&	III		\\
WBIS\_053920.4-022736.8	&	M2	&	3490	&	-0.42	&	0.30	&	III	&	-	&	-2.56	&	III		\\
WBIS\_053925.3-023143.6	&	G3.5	&	5680	&	-0.33	&	1.0	&	III	&	-	&	-2.76	&	III		\\
WBIS\_053907.6-022823.3	&	M3	&	3360	&	-0.77	&	0.30	&	III	&	-	&	-2.48	&	III		\\
WBIS\_053922.9-023333.0	&	M2	&	3490	&	-0.73	&	0.30	&	III	&	-	&	-2.40	&	III		\\
WBIS\_053914.5-022833.3	&	M3.5	&	3260	&	-0.91	&	0.20	&	III	&	-	&	-2.12	&	III		\\
WBIS\_053908.2-023228.4	&	M5	&	2880	&	-1.13	&	0.09	&	III	&	-	&	-2.02	&	III		\\
WBIS\_053859.5-024508.0	&	F3	&	6660	&	0.37	&	1.4	&	-	&	-	&	-2.90	&	III	\\
WBIS\_053913.5-023739.0	&	M4	&	3160	&	-0.98	&	0.17	&	III	&	-	&	-1.97	&	III		\\
WBIS\_053915.8-023650.7	&	M4	&	3160	&	-0.93	&	0.17	&	II	&	II	&	-1.61	&	III		\\
WBIS\_053926.5-022615.4	&	M1.5	&	3560	&	-0.90	&	0.40	&	III	&	-	&	-2.20	&	III		\\
WBIS\_053823.1-023649.3	&	M4.5	&	3020	&	-1.10	&	0.13	&	II	&	II	&	-1.75	&	III	\\
\hline
\end{tabular}
\end{table}
\end{landscape}

\begin{landscape}
\begin{table}[htb]
\tabularfont
\begin{tabular}{lcccccccc}
\topline
Object ID & SpT & T$_\mathrm{eff}$ & log(L$_\mathrm{bol}$/L$_\mathrm{\odot}$) & Mass & Class (H07)  & Class (K15) & $\alpha$ & Class (This work) \\
& & (K) & & (M$_\mathrm{\odot}$) & & & & \\ \midline
WBIS\_053834.6-024108.8	&	M2	&	3490	&	-0.82	&	0.40	&	II	&	II	&	-1.70	&	III		\\
WBIS\_053834.3-023500.1	&	K5	&	4140	&	0.04	&	0.80	&	III	&	-	&	-2.84	&	III		\\
WBIS\_053829.1-023602.7	&	M1.5	&	3560	&	-0.67	&	0.40	&	III	&	-	&	-2.69	&	III	\\
WBIS\_053845.3-023729.3	&	M5.5	&	2920	&	-0.99	&	0.11	&	II	&	-	&	-1.80	&	III		\\
WBIS\_053849.9-024122.8	&	M3	&	3360	&	-0.78	&	0.30	&	III	&	-	&	-2.48	&	III		\\
WBIS\_053920.2-023825.9	&	M5.5	&	2920	&	-1.07	&	0.10	&	III	&	-	&	-2.42	&	III		\\
WBIS\_053826.2-024041.3	&	M4.5	&	3020	&	-1.53	&	0.10	&	III	&	-	&	-2.46	&	III		\\
WBIS\_053814.7-024015.2	&	L1	&	2102	&	-2.42	&	0.007	&	-	&	-	&	-2.56	&	III		\\
WBIS\_053915.8-023826.3	&	M4.5	&	3020	&	-1.58	&	0.10	&	-	&	-	&	-2.23	&	III		\\
WBIS\_053911.8-022741.0	&	M4.5	&	3020	&	-1.03	&	0.13	&	III	&	-	&	-2.73	&	III		\\
WBIS\_053851.5-023620.6	&	K5	&	4140	&	-0.60	&	0.80	&	EV	&	-	&	-2.56	&	III		\\
WBIS\_053825.7-023121.7	&	M3.5	&	3260	&	-1.42	&	0.20	&	III	&	-	&	-2.51	&	III		\\
WBIS\_053805.5-023557.1	&	M4.5	&	3020	&	-1.73	&	0.09	&	II	&	-	&	-2.03	&	III		\\
WBIS\_053849.2-023822.3	&	K7	&	3970	&	-0.05	&	0.60	&	III	&	-	&	-2.85	&	III		\\
WBIS\_053817.4-024024.2	&	M6	&	2860	&	-1.56	&	0.06	&	-	&	-	&	-2.32	&	III		\\
WBIS\_053829.6-022514.2	&	M4	&	3160	&	-1.51	&	0.15	&	III	&	-	&	-2.43	&	III		\\
WBIS\_053848.7-023616.3	&	M1.5	&	3560	&	-0.60	&	0.40	&	III	&	-	&	-2.91	&	III		\\
WBIS\_053850.0-023735.5	&	M5.5	&	2920	&	-0.89	&	0.11	&	III	&	-	&	-2.55	&	III		\\
WBIS\_053846.0-024523.1	&	M5	&	2880	&	-1.05	&	0.10	&	III	&	-	&	-2.54	&	III		\\
WBIS\_053823.3-024414.2	&	M4	&	3160	&	-0.99	&	0.17	&	III	&	-	&	-2.72	&	III		\\
WBIS\_053821.4-023336.3	&	M6	&	2860	&	-1.77	&	0.06	&	III	&	-	&	-2.49	&	III		\\
WBIS\_053823.5-024131.7	&	M4.5	&	3020	&	-0.95	&	0.13	&	III	&	-	&	-2.68	&	III	\\
WBIS\_053818.3-023538.6	&	M7	&	2683	&	-1.67	&	0.04	&	III	&	-	&	-2.26	&	III		\\
WBIS\_053852.6-023215.5	&	M6.5	&	2815	&	-2.14	&	0.04	&	-	&	-	&	-2.44	&	III		\\
WBIS\_053820.2-023801.6	&	M4.5	&	3020	&	-0.71	&	0.15	&	III	&	-	&	-2.64	&	III		\\
WBIS\_053813.2-022407.5	&	M5.5	&	2920	&	-1.28	&	0.09	&	III	&	-	&	-2.47	&	III		\\
WBIS\_053853.8-024458.8	&	M6	&	2860	&	-1.83	&	0.05	&	-	&	-	&	-2.49	&	III \\
WBIS\_053911.4-023332.8	&	M5	&	2880	&	-1.41	&	0.07	&	III	&	-	&	-2.53	&	III		\\
WBIS\_053849.7-023452.7	&	M5	&	2880	&	-0.88	&	0.11	&	II	&	II	&	-2.08	&	III		\\
WBIS\_053850.8-023626.8	&	M3	&	3360	&	-0.88	&	0.30	&	III	&	-	&	-2.47	&	III		\\
WBIS\_053908.9-023957.9	&	M7	&	2683	&	-1.43	&	0.05	&	III	&	-	&	-2.29	&	III \\
WBIS\_053838.6-024155.9	&	M5.5	&	2920	&	-1.44	&	0.075	&	III	&	-	&	-2.50	&	III		\\
WBIS\_053851.7-023603.4	&	M6	&	2860	&	-0.87	&	0.11	&	III	&	-	&	-2.52	&	III		\\
WBIS\_053836.9-023643.3	&	M4.5	&	3020	&	-0.88	&	0.13	&	III	&	-	&	-2.62	&	III		\\
WBIS\_053847.2-023540.6	&	B2	&	20600	&	3.43	&	7.30	&	III	&	-	&	-3.03	&	III		\\
WBIS\_053904.5-023835.3	&	M6	&	2860	&	-1.53	&	0.06	&	III	&	-	&	-2.37	&	III		\\
WBIS\_053908.1-022844.8	&	M4	&	3160	&	-1.22	&	0.17	&	EV	&	-	&	-2.43	&	III		\\
WBIS\_053844.8-023600.2	&	O9.5	&	31900	&	4.72	&	18.70	&	-	&	-	&	-	&	-	\\
\hline
\end{tabular}
\end{table}
\end{landscape}

\begin{landscape}
\begin{table}[htb]
\tabularfont
\begin{tabular}{lcccccccc}
\topline
Object ID & SpT & T$_\mathrm{eff}$ & log(L$_\mathrm{bol}$/L$_\mathrm{\odot}$) & Mass & Class (H07)  & Class (K15) & $\alpha$ & Class (This work) \\
& & (K) & & (M$_\mathrm{\odot}$) & & & & \\ \midline
WBIS\_053841.5-023552.3	&	M6	&	2860	&	-1.25	&	0.08	&	-	&	-	&	-	&	-		\\
WBIS\_053844.1-023606.3	&	A2	&	8800	&	1.38	&	1.98	&	-	&	-	&	-	&	-	\\
WBIS\_053845.3-023541.2	&	M5.5	&	2920	&	-1.09	&	0.10	&	-	&	-	&	-	&	-		\\
WBIS\_053835.3-023313.1	&	M5.5	&	2920	&	-1.21	&	0.09	&	-	&	-	&	-	&	-		\\
WBIS\_053835.8-023313.4	&	M5.5	&	2920	&	-1.14	&	0.10	&	-	&	-	&	-	&	-	\\
WBIS\_053842.4-023604.5	&	M6	&	2860	&	-1.30	&	0.075	&	-	&	-	&	-	&	-		\\
WBIS\_053844.5-024037.7	&	M6.5	&	2815	&	-1.54	&	0.05	&	-	&	-	&	-	&	-		\\
WBIS\_053913.8-023145.6	&	M8.5	&	2467	&	-2.57	&	0.015	&	-	&	-	&	-	&	-		\\
WBIS\_053910.8-023714.6	&	M9	&	2394	&	-2.61	&	0.01	&	-	&	-	&	-	&	-		\\
WBIS\_053903.6-022536.7	&	L0	&	2248	&	-2.93	&	0.01	&	-	&	-	&	-	&	-		\\
WBIS\_053852.7-022843.7	&	L0	&	2248	&	-3.47	&	0.04	&	-	&	-	&	-	&	-	\\
WBIS\_053829.5-022937.0	&	L1	&	2102	&	-3.01	&	0.008	&	-	&	-	&	-	&	-		\\
WBIS\_053857.5-022905.5	&	L1	&	2102	&	-3.08	&	0.009	&	-	&	-	&	-	&	-	\\
WBIS\_053803.2-022656.7	&	L1	&	2102	&	-3.20	&	0.01	&	-	&	-	&	-	&	-		\\
WBIS\_053826.1-022305.0	&	L3.5	&	1756	&	-3.41	&	0.005	&	-	&	-	&	-	&	-		\\
WBIS\_053839.2-022805.8	&	L5	&	1581	&	-3.20	&	0.004	&	-	&	-	&	-	&	-		\\
WBIS\_053844.5-024030.5	&	-	&	-	&	-	&	-	&	-	&	-	&	-	&	-	\\
WBIS\_053848.3-023641.0	&	M4.5	&	3020	&	-0.63	&	0.15	&	-	&	-	&	-	&	-	\\
WBIS\_053841.4-023644.5	&	M2	&	3490	&	-0.81	&	0.40	&	-	&	-	&	-	&	-		\\
WBIS\_053847.5-023524.9	&	M1	&	3630	&	-0.23	&	0.40	&	-	&	-	&	-	&	-		\\
WBIS\_053846.8-023643.6	&	M5	&	2880	&	-0.98	&	0.10	&	-	&	-	&	-	&	-		\\
WBIS\_053843.0-023614.6	&	M1	&	3630	&	-0.30	&	0.40	&	-	&	-	&	-	&	-		\\
WBIS\_053845.6-023559.0	&	B2	&	20600	&	3.43	&	7.30	&	-	&	-	&	-	&	-		\\
WBIS\_053903.2-023020.0	&	M9	&	2394	&	-2.47	&	0.01	&	-	&	-	&	-	&	-		\\
\hline
\end{tabular}
\end{table}
\end{landscape}

\vspace{-3em}

\section*{Acknowledgements}
The authors thank the anonymous referee for the constructive report which has helped improve the overall quality of the paper. Based on observations obtained with WIRCam, a joint project of CFHT, Taiwan, Korea, Canada, and France, at the
Canada–France–Hawaii Telescope (CFHT) which is operated by the National Research Council (NRC) of Canada, the Institut National des Sciences de l’Univers of the Centre National de la Recherche Scientifique of France, and the University of Hawaii. This work has made use of data from the European Space Agency (ESA) mission
{\it Gaia} (\url{https://www.cosmos.esa.int/gaia}), processed by the {\it Gaia} Data Processing and Analysis Consortium (DPAC, \url{https://www.cosmos.esa.int/web/gaia/dpac/consortium}). Funding for the DPAC has been provided by national institutions, in particular the institutions participating in the {\it Gaia} Multilateral Agreement. This publication makes use of data products from the Two Micron All Sky Survey, which is a joint project of the University of Massachusetts and the Infrared Processing and Analysis Center/California Institute of Technology, funded by the National Aeronautics and Space Administration and the National Science Foundation. This publication makes use of data products from the Wide-field Infrared Survey Explorer, which is a joint project of the University of California, Los Angeles, and the Jet Propulsion Laboratory/California Institute of Technology, funded by the National Aeronautics and Space Administration. This work is based [in part] on observations made with the Spitzer Space Telescope, which was operated by the Jet Propulsion Laboratory, California Institute of Technology under a contract with NASA. BD is thankful to the Center for Research, CHRIST (Deemed to be University), Bangalore, India. JJ acknowledges
the financial support received through the DST-SERB
grant SPG/2021/003850.

\vspace{-1em}




\begin{theunbibliography}{}
\vspace{-1.5em}

\bibitem[\protect\citeauthoryear{Allers \& Liu}{2020}]{allers2020} Allers K.~N., Liu M.~C., 2020, PASP, 132, 104401. 

\bibitem[\protect\citeauthoryear{Andre, Ward-Thompson, \& Barsony}{2000}]{andre2000} Andre P., Ward-Thompson D., Barsony M., 2000, prpl.conf, 59

\bibitem[\protect\citeauthoryear{Andre, Ward-Thompson, \& Greaves}{2012}]{andre2012} Andr{\'e} P., Ward-Thompson D., Greaves J., 2012, Sci, 337, 69. 


\bibitem[\protect\citeauthoryear{Andrews}{2020}]{andrews2020} Andrews S.~M., 2020, ARA\&A, 58, 483. 

\bibitem[\protect\citeauthoryear{Ansdell \textit{et al}.}{2017}]{ansdell2017} Ansdell M., Williams J.~P., Manara C.~F., Miotello A., Facchini S., van der Marel N., Testi L., et al., 2017, AJ, 153, 240. 

\bibitem[\protect\citeauthoryear{Baraffe \textit{et al.}}{2003}]{baraffe2003} Baraffe I., Chabrier G., Barman T.~S., Allard F., Hauschildt P.~H., 2003, A\&A, 402, 701. 

\bibitem[\protect\citeauthoryear{Baraffe \textit{et al.}}{2015}]{baraffe2015} Baraffe I., Homeier D., Allard F., Chabrier G., 2015, A\&A, 577, A42. 

\bibitem[\protect\citeauthoryear{B{\'e}jar, Zapatero Osorio, \& Rebolo}{2004}]{bejar2004} B{\'e}jar V.~J.~S., Zapatero Osorio M.~R., Rebolo R., 2004, AN, 325, 705. 

\bibitem[\protect\citeauthoryear{Bouvier \textit{et al}.}{2007}]{bouvier2007} Bouvier J., Alencar S.~H.~P., Harries T.~J., Johns-Krull C.~M., Romanova M.~M., 2007, prpl.conf, 479

\bibitem[\protect\citeauthoryear{Caballero \textit{et al.}}{2007}]{caballero2007} Caballero J.~A., B{\'e}jar V.~J.~S., Rebolo R., Eisl{\"o}ffel J., Zapatero Osorio M.~R., Mundt R., Barrado Y Navascu{\'e}s D., et al., 2007, A\&A, 470, 903. 

\bibitem[\protect\citeauthoryear{Calvet \textit{et al}.}{2004}]{calvet2004} Calvet N., Muzerolle J., Brice{\~n}o C., Hern{\'a}ndez J., Hartmann L., Saucedo J.~L., Gordon K.~D., 2004, AJ, 128, 1294.

\bibitem[\protect\citeauthoryear{Cardelli, Clayton, \& Mathis}{1989}]{cardelli1989} Cardelli J.~A., Clayton G.~C., Mathis J.~S., 1989, ApJ, 345, 245.

\bibitem[\protect\citeauthoryear{Currie \textit{et al.}}{2009}]{currie2009} Currie T., Lada C.~J., Plavchan P., Robitaille T.~P., Irwin J., Kenyon S.~J., 2009, ApJ, 698, 1.

\bibitem[\protect\citeauthoryear{Cutri \textit{et al.}}{2003}]{cutri2003} Cutri R.~M., Skrutskie M.~F., van Dyk S., Beichman C.~A., Carpenter J.~M., Chester T., Cambresy L., et al., 2003, yCat, II/246

\bibitem[\protect\citeauthoryear{Cutri \& \textit{et al.}}{2012}]{cutri2012} Cutri R.~M., et al., 2012, yCat, II/311

\bibitem[\protect\citeauthoryear{Dahm \& Hillenbrand}{2007}]{dahm2007} Dahm S.~E., Hillenbrand L.~A., 2007, AJ, 133, 2072.

\bibitem[\protect\citeauthoryear{Damian \textit{et al.}}{2021}]{damian2021} Damian B., Jose J., Samal M.~R., Moraux E., Das S.~R., Patra S., 2021, MNRAS, 504, 2557. 

\bibitem[\protect\citeauthoryear{Damian \textit{et al.}}{2023}]{damian2023} Damian B., Jose J., Biller B., Herczeg, G. J., Albert, L., Allers, K. N., Zhang, Z., et al., 2023, arXiv, arXiv:2303.17424.

\bibitem[\protect\citeauthoryear{Das \textit{et al}.}{2021}]{das2021} Das S.~R., Jose J., Samal M.~R., Zhang S., Panwar N., 2021, MNRAS, 500, 3123. 

\bibitem[\protect\citeauthoryear{Das \textit{et al.}}{2023}]{das2023} Das S.~R., Gupta S., Prakash P., Samal M., Jose J., 2023, ApJ, 948, 7. 

\bibitem[\protect\citeauthoryear{Dubber \textit{et al.}}{2021}]{dubber2021} Dubber S., Biller B., Allers K., Jose J., Albert L., Pantoja B., Fontanive C., et al., 2021, MNRAS, 505, 4215.

\bibitem[\protect\citeauthoryear{Esplin \& Luhman}{2017}]{esplin2017} Esplin T.~L., Luhman K.~L., 2017, AJ, 154, 134. 

\bibitem[\protect\citeauthoryear{Esplin \& Luhman}{2022}]{esplin2022} Esplin T.~L., Luhman K.~L., 2022, AJ, 163, 64. 

\bibitem[\protect\citeauthoryear{Fabricius \textit{et al.}}{2021}]{fabricius2021} Fabricius C., Luri X., Arenou F., Babusiaux C., Helmi A., Muraveva T., Reyl{\'e} C., et al., 2021, A\&A, 649, A5. 

\bibitem[\protect\citeauthoryear{Filippazzo \textit{et al.}}{2015}]{filippazzo2015} Filippazzo J.~C., Rice E.~L., Faherty J., Cruz K.~L., Van Gordon M.~M., Looper D.~L., 2015, ApJ, 810, 158. 

\bibitem[\protect\citeauthoryear{Frank \textit{et al}.}{2014}]{frank2014} Frank A., Ray T.~P., Cabrit S., Hartigan P., Arce H.~G., Bacciotti F., Bally J., et al., 2014, prpl.conf, 451. 

\bibitem[\protect\citeauthoryear{Furlan \textit{et al.}}{2006}]{furlan2006} Furlan E., Hartmann L., Calvet N., D'Alessio P., Franco-Hern{\'a}ndez R., Forrest W.~J., Watson D.~M., et al., 2006, ApJS, 165, 568. 

\bibitem[\protect\citeauthoryear{Gaia Collaboration}{2022}]{gaia2022} Gaia Collaboration, 2022, yCat, I/355

\bibitem[\protect\citeauthoryear{Greene \textit{et al.}}{1994}]{greene1994} Greene T.~P., Wilking B.~A., Andre P., Young E.~T., Lada C.~J., 1994, ApJ, 434, 614. 

\bibitem[\protect\citeauthoryear{Gupta\textit{ et al.}}{2021}]{gupta2021} Gupta S., Jose J., More S., Das S.~R., Herczeg G.~J., Samal M.~R., Guo Z., \textit{et al.}, 2021, MNRAS, 508, 3388. 

\bibitem[\protect\citeauthoryear{Guzm{\'a}n-D{\'\i}az \textit{et al.}}{2021}]{guzman2021} Guzm{\'a}n-D{\'\i}az J., Mendigut{\'\i}a I., Montesinos B., Oudmaijer R.~D., Vioque M., Rodrigo C., Solano E., et al., 2021, A\&A, 650, A182. 

\bibitem[\protect\citeauthoryear{Han\textit{ et al.}}{2013}]{han2013} Han C., Jung Y.~K., Udalski A., Sumi T., Gaudi B.~S., Gould A., Bennett D.~P., et al., 2013, ApJ, 778, 38.


\bibitem[\protect\citeauthoryear{Hartigan, Edwards, \& Ghandour}{1995}]{hartigan1995} Hartigan P., Edwards S., Ghandour L., 1995, ApJ, 452, 736. 

\bibitem[\protect\citeauthoryear{Hartmann}{2009}]{hartmann2009} Hartmann L., 2009, apsf.book

\bibitem[\protect\citeauthoryear{Hartmann, Herczeg, \& Calvet}{2016}]{hartmann2016} Hartmann L., Herczeg G., Calvet N., 2016, ARA\&A, 54, 135. 

\bibitem[\protect\citeauthoryear{Herczeg \& Hillenbrand}{2014}]{herczeg2014} Herczeg G.~J., Hillenbrand L.~A., 2014, ApJ, 786, 97. 

\bibitem[\protect\citeauthoryear{Hern{\'a}ndez \textit{et al}.}{2007}]{hernandez2007int} Hern{\'a}ndez J., Calvet N., Brice{\~n}o C., Hartmann L., Vivas A.~K., Muzerolle J., Downes J., et al., 2007, ApJ, 671, 1784. 

\bibitem[\protect\citeauthoryear{Hern{\'a}ndez \textit{et al}.}{2007}]{hernandez2007} Hern{\'a}ndez J., Hartmann L., Megeath T., Gutermuth R., Muzerolle J., Calvet N., Vivas A.~K., et al., 2007, ApJ, 662, 1067.

\bibitem[\protect\citeauthoryear{Hillenbrand \textit{et al}.}{1998}]{hillenbrand1998} Hillenbrand L.~A., Strom S.~E., Calvet N., Merrill K.~M., Gatley I., Makidon R.~B., Meyer M.~R., et al., 1998, AJ, 116, 1816. 


\bibitem[\protect\citeauthoryear{Johnson \textit{et al.}}{2010}]{johnson2010} Johnson J.~A., Aller K.~M., Howard A.~W., Crepp J.~R., 2010, PASP, 122, 905. 

\bibitem[\protect\citeauthoryear{Jose \textit{et al.}}{2020}]{jose2020} Jose J., Biller B.~A., Albert L., Dubber S., Allers K., Herczeg G.~J., Liu M.~C., et al., 2020, ApJ, 892, 122. 

\bibitem[\protect\citeauthoryear{Koenig \& Leisawitz}{2014}]{koenig2014} Koenig X.~P., Leisawitz D.~T., 2014, ApJ, 791, 131. 

\bibitem[\protect\citeauthoryear{Koenig \textit{et al.}}{2015}]{koenig2015} Koenig X., Hillenbrand L.~A., Padgett D.~L., DeFelippis D., 2015, AJ, 150, 100.

\bibitem[\protect\citeauthoryear{Kordopatis \textit{et al.}}{2023}]{kordopatis2023} Kordopatis G., Schultheis M., McMillan P.~J., Palicio P.~A., de Laverny P., Recio-Blanco A., Creevey O., et al., 2023, A\&A, 669, A104. 

\bibitem[\protect\citeauthoryear{Lada}{1987}]{lada1987} Lada C.~J., 1987, IAUS, 115, 1

\bibitem[\protect\citeauthoryear{Lada \& Lada}{1995}]{lada1995} Lada E.~A., Lada C.~J., 1995, AJ, 109, 1682. 

\bibitem[\protect\citeauthoryear{Lada \textit{et al}.}{2006}]{lada2006} Lada C.~J., Muench A.~A., Luhman K.~L., Allen L., Hartmann L., Megeath T., Myers P., et al., 2006, AJ, 131, 1574. 

\bibitem[\protect\citeauthoryear{Lalchand \textit{et al.}}{2022}]{lalchand2022} Lalchand B., Chen W.-P., Biller B.~A., Albert L., Allers K., Dubber S., Zhang Z., et al., 2022, AJ, 164, 125. 

\bibitem[\protect\citeauthoryear{Luhman \textit{et al.}}{2005}]{luhman2005} Luhman K.~L., Lada C.~J., Hartmann L., Muench A.~A., Megeath S.~T., Allen L.~E., Myers P.~C., et al., 2005, ApJL, 631, L69. 

\bibitem[\protect\citeauthoryear{Luhman\textit{ et al.}}{2008}]{luhman2008} Luhman K.~L., Hern{\'a}ndez J., Downes J.~J., Hartmann L., Brice{\~n}o C., 2008, ApJ, 688, 362. 

\bibitem[\protect\citeauthoryear{Luhman}{2012}]{luhman2012} Luhman K.~L., 2012, ARA\&A, 50, 65. 

\bibitem[\protect\citeauthoryear{Manara \textit{et al.}}{2021}]{manara2021} Manara C.~F., Frasca A., Venuti L., Siwak M., Herczeg G.~J., Calvet N., Hernandez J., et al., 2021, A\&A, 650, A196. 

\bibitem[\protect\citeauthoryear{Meyer \textit{et al}.}{2007}]{meyer2007} Meyer M.~R., Backman D.~E., Weinberger A.~J., Wyatt M.~C., 2007, prpl.conf, 573

\bibitem[\protect\citeauthoryear{Michel, van der Marel, \& Matthews}{2021}]{michel2021} Michel A., van der Marel N., Matthews B.~C., 2021, ApJ, 921, 72. 

\bibitem[\protect\citeauthoryear{Miotello \textit{et al}.}{2022}]{miotello2022} Miotello A., Kamp I., Birnstiel T., Cleeves L.~I., Kataoka A., 2022, arXiv, arXiv:2203.09818

\bibitem[\protect\citeauthoryear{Monteiro \textit{et al.}}{2020}]{monteiro2020} Monteiro H., Dias W.~S., Moitinho A., Cantat-Gaudin T., L{\'e}pine J.~R.~D., Carraro G., Paunzen E., 2020, MNRAS, 499, 1874. 

\bibitem[\protect\citeauthoryear{Muench \textit{et al.}}{2001}]{muench2001} Muench A.~A., Alves J., Lada C.~J., Lada E.~A., 2001, ApJL, 558, L51. 

\bibitem[\protect\citeauthoryear{Natta \textit{et al.}}{2002}]{natta2002} Natta A., Testi L., Comer{\'o}n F., Oliva E., D'Antona F., Baffa C., Comoretto G., et al., 2002, A\&A, 393, 597. 

\bibitem[\protect\citeauthoryear{Padoan \& Nordlund}{2002}]{padoan2002} Padoan P., Nordlund {\r{A}}., 2002, ApJ, 576, 870.

\bibitem[\protect\citeauthoryear{Panwar \textit{et al.}}{2017}]{panwar2017} Panwar N., Samal M.~R., Pandey A.~K., Jose J., Chen W.~P., Ojha D.~K., Ogura K., et al., 2017, MNRAS, 468, 2684. 

\bibitem[\protect\citeauthoryear{Pascucci \textit{et al}.}{2022}]{pascucci2022} Pascucci I., Cabrit S., Edwards S., Gorti U., Gressel O., Suzuki T., 2022, arXiv, arXiv:2203.10068

\bibitem[\protect\citeauthoryear{Pecaut \& Mamajek}{2013}]{pecaut2013} Pecaut M.~J., Mamajek E.~E., 2013, ApJS, 208, 9. 

\bibitem[\protect\citeauthoryear{Pe{\~n}a Ram{\'\i}rez \textit{et al.}}{2012}]{pena2012} Pe{\~n}a Ram{\'\i}rez K., B{\'e}jar V.~J.~S., Zapatero Osorio M.~R., Petr-Gotzens M.~G., Mart{\'\i}n E.~L., 2012, ApJ, 754, 30. 

\bibitem[\protect\citeauthoryear{Penoyre, Belokurov, \& Evans}{2022}]{penoyre2022} Penoyre Z., Belokurov V., Evans N.~W., 2022, MNRAS, 513, 5270. 

\bibitem[\protect\citeauthoryear{Puget \textit{et al.}}{2004}]{puget2004} Puget P., Stadler E., Doyon R., Gigan P., Thibault S., Luppino G., Barrick G., et al., 2004, SPIE, 5492, 978.

\bibitem[\protect\citeauthoryear{Rayner \textit{et al.}}{2003}]{rayner2003} Rayner J.~T., Toomey D.~W., Onaka P.~M., Denault A.~J., Stahlberger W.~E., Vacca W.~D., Cushing M.~C., et al., 2003, PASP, 115, 362.

\bibitem[\protect\citeauthoryear{Reipurth \& Clarke}{2001}]{reipurth2001} Reipurth B., Clarke C., 2001, AJ, 122, 432.

\bibitem[\protect\citeauthoryear{Ribas, Bouy, \& Mer{\'\i}n}{2015}]{ribas2015} Ribas {\'A}., Bouy H., Mer{\'\i}n B., 2015, A\&A, 576, A52.


\bibitem[\protect\citeauthoryear{Rilinger \& Espaillat}{2021}]{rilinger2021} Rilinger A.~M., Espaillat C.~C., 2021, ApJ, 921, 182. 


\bibitem[\protect\citeauthoryear{Sherry \textit{et al}.}{2008}]{sherry2008} Sherry W.~H., Walter F.~M., Wolk S.~J., Adams N.~R., 2008, AJ, 135, 1616. 

\bibitem[\protect\citeauthoryear{Scholz \& Jayawardhana}{2008}]{scholz2008} Scholz A., Jayawardhana R., 2008, ApJL, 672, L49. 

\bibitem[\protect\citeauthoryear{Scholz \textit{et al.}}{2023}]{scholz2023} Scholz A., Muzic K., Jayawardhana R., Almendros-Abad V., Wilson I., 2023, arXiv, arXiv:2303.12451.

\bibitem[\protect\citeauthoryear{Shvartzvald \textit{et al.}}{2017}]{shvartzvald2017} Shvartzvald Y., Yee J.~C., Calchi Novati S., Gould A., Lee C.-U., Beichman C., Bryden G., et al., 2017, ApJL, 840, L3. 

\bibitem[\protect\citeauthoryear{Spitzer Science Center (SSC) \& Infrared Science Archive (IRSA)}{2021}]{spitzer2021} Spitzer Science Center (SSC), Infrared Science Archive (IRSA), 2021, yCat, II/368

\bibitem[\protect\citeauthoryear{Stamatellos \textit{et al}.}{2011}]{stamatellos2011} Stamatellos D., Maury A., Whitworth A., Andr{\'e} P., 2011, MNRAS, 413, 1787.

\bibitem[\protect\citeauthoryear{Stoop \textit{et al.}}{2023}]{stoop2023} Stoop M., Kaper L., de Koter A., Guo D., Lamers H.~J.~G.~L.~M., Rieder S., 2023, A\&A, 670, A108.

\bibitem[\protect\citeauthoryear{Wang \& Chen}{2019}]{wang2019} Wang S., Chen X., 2019, ApJ, 877, 116. 

\bibitem[\protect\citeauthoryear{Williams \& Cieza}{2011}]{williams2011} Williams J.~P., Cieza L.~A., 2011, ARA\&A, 49, 67. 

\bibitem[\protect\citeauthoryear{Winter\textit{ et al}.}{2018}]{winter2018} Winter A.~J., Clarke C.~J., Rosotti G., Ih J., Facchini S., Haworth T.~J., 2018, MNRAS, 478, 2700. 

\bibitem[\protect\citeauthoryear{Whitworth \& Zinnecker}{2004}]{whitworth2004} Whitworth A.~P., Zinnecker H., 2004, A\&A, 427, 299. 

\bibitem[\protect\citeauthoryear{Yao \textit{et al}.}{2018}]{yuhan2018} Yao Y., Meyer M.~R., Covey K.~R., Tan J.~C., Da Rio N., 2018, ApJ, 869, 72. 


\end{theunbibliography}
\end{document}